\documentclass[aps,prl,twocolumn,groupedaddress]{revtex4-1}
\usepackage{graphicx}
\usepackage{epsfig}
\usepackage{amsfonts}
\usepackage{amssymb}

\begin{document}

\title{Two distinct electronic sites in the Cu-O plane of the La$_{2-x}$Sr$_{x}$CuO$_{4}$ pseudogap state}

\author{R.X. Smith}
\email{robsmith@magnet.fsu.edu}
\author{P.L. Kuhns and A.P. Reyes and G.S. Boebinger}

\affiliation{National High Magnetic Field Laboratory, Florida State University, Tallahassee, FL 32310}

\date{\today}

\begin{abstract}
The pseudogap state is widely regarded as a precursor to high-temperature superconductivity yet remains poorly understood. Using high field (30 T) NMR, we report two electronically distinct oxygen sites within the copper-oxygen (Cu-O) planes of an underdoped cuprate in the pseudogap state. At temperatures well above the bulk superconducting transition, roughly 25$\%$ of the oxygen sites evidence superconducting pair formation, the remaining evidence antiferromagnetic ordering on the nearest neighbor Cu sites. Any superconducting fluctuations are slower than the microsecond timescales of the NMR probe, a confirmation that the (nearly) static bimodality of electronic sites in the Cu-O plane - first reported using scanning tunneling microscopy on surfaces - extends throughout the bulk pseudogap state.
\end{abstract}


\maketitle
The two dimensional physics of the copper-oxygen (Cu-O) planes in the cuprate high temperature superconductors (HTS) is widely thought to govern their superconductivity.   Focus has been given to the pseudogap state, the precursor state to superconductivity for underdoped samples~\cite{KivelsonRMP75, OrensteinSCIENCE288, NormanAdvPhys54}.  Nernst effect measurements on underdoped La$_{2-x}$Sr$_{x}$CuO$_{4}$ (LSCO) evidence superconducting fluctuations in the pseudogap state well above the superconducting transition temperature, T$_C$~\cite{XuNAT406}, a temperature regime that overlaps with the strong magnetic fluctuations seen in neutron scattering  \cite{ThurstonPRB40, TranquadaNAT375} and nuclear quadrupole resonance (NQR) experiments \cite{HuntPRL82}.  Scanning tunneling microscopy (STM) on the surface of BSCCO, another high temperature superconductor, has revealed with atomic resolution that the local electronic density of states is patterned on a nanometer length scale.  This patterning shows a bimodal distribution -either very high or very low electron density- on each oxygen site in the Cu-O plane \cite{KohsakaSCI315}.  In this Letter, we report oxygen NMR measurements that probe the local electron density on the planar oxygen sites in LSCO, finding that bimodal electron density is a bulk property of the pseudogap state and, hence, underlies the phenomena reported in the Nernst and neutron scattering experiments.

There is a long history of NMR studies of the HTS materials in magnetic fields up to 9 T that have found low frequency spin and charge fluctuations in the Cu-O planes that has been discussed in terms of the formation of nano-scale phase separation \cite{AlloulPRL63, SongPRL70, OhsugiJPSJ63, WalstedtPRL72, HuntPRL82, CurroPRL85, SingerPRB72}, although the evidence is indirect.  In this Letter, we find that high magnetic fields of 30 T increase the resolution of the NMR spectra to the extent that we can directly observe for the first time two distinct electronic environments on the oxygen atoms in the Cu-O plane.  We study oxygen NMR because (A) $^{17}$O NMR has shown that doped holes reside primarily on the planar oxygen sites \cite{HaasePRB69, ZhengJPSJ64, HammelPRB57} and  (B) copper NMR in underdoped La$_{2-x}$Sr$_{x}$CuO$_{4}$ undergoes a 'wipeout'- a dramatic reduction- of the $^{63}$Cu NMR signal below $\sim$70K due to the inhomogeneous slowing of strong spin-fluctuations \cite{HuntPRL82}.  There is another benefit of the 30 T magnetic fields: the superconducting transition temperature is suppressed from T$_C$ $\sim$30 K to $\sim$4 K, enabling high-field NMR experiments to probe the bulk properties of the pseudogap state to low temperatures.

La$_{2-x}$Sr$_{x}$CuO$_{4}$ is studied in the underdoped regime, x=0.115, using NMR as a bulk probe of local magnetic field and local electric field gradient (EFG) at the planar oxygen sites. The sample is a single crystal grown in a floating-zone furnace.  The superconducting transition, T$_{C}$ = 30 K in zero magnetic field, has a width of $\sim$2.5 K.  The 30 T magnetic field suppresses T$_{C}$ to $\sim$4 K.  Due to the low natural abundance of $^{17}$O, the sample was annealed at 900 C in a $^{17}$O enriched gas for $\sim$100 hours.

NMR spectra were recorded in the range 4-300 K following field cooling in applied magnetic fields of H$_{0}$ = 8 T, 14 T and 30 T using standard spin echo techniques at fixed frequency while sweeping the magnetic field. For La$_{2-x}$Sr$_{x}$CuO$_{4}$, the $^{17}$O (I=5/2) nuclear levels are perturbed by the interaction of the quadrupole moment with an EFG.  The spectra show five peaks due to the five nonequivalent transitions between the perturbed levels.  Each occurs at a resonant field $^{17}\gamma$H$_{n}$=($\nu_{L}$-n$\nu_{c}$)/(1+$^{17}$K) where $\nu_{L}$ is a fixed reference frequency, $\nu_{c}$ is the quadrupolar splitting when H$_{0}\|$c, $^{17}$K is the magnetic shift and n = (-2,-1,0,1,2) denumerates the nuclear energy level transitions.  The Zeeman splitting of the nuclear energy levels, $\nu_{L}$, is 46.1752 MHz, 80.8066 MHz and 173.157 MHz at 8 T, 14 T and 30 T, respectively, while the quadrupolar interaction yields a spacing between transition peaks of $\nu_{c}\sim$200 kHz, consistent with previous reports \cite{HaasePhysC341, HaaseJoS15}.  The spectral shape is independent of the sample cooling protocol in either high field or zero field.

\begin{figure}[t]
\begin{center}
\includegraphics[width=8cm]{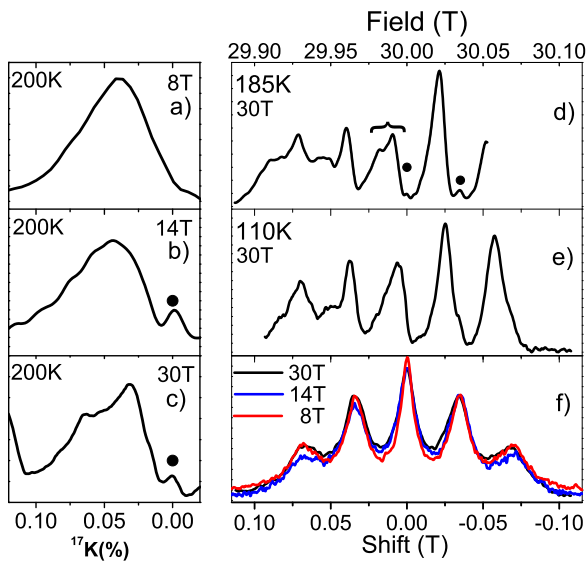}
\end{center}
\caption{Central transition (CT) of the planar $^{17}$O NMR spectrum at 200 K for applied fields of a) 8 T, b) 14 T, and c) 30 T, expressed as Knight shift. A line splitting becomes visible at 30T. Dots indicate the apical CT, where visible, that has been suppressed as described in the text. (d)-(f) Evolution of the NMR 30 T field-swept spectra with decreasing temperature. Shift is defined from the peak position of the CT in the superconducting state. Panel (f) includes spectra taken at 8 T and 14 T, showing that the spectral shape and linewidth has become independent of applied field at low temperatures.
\label{fig:LSCuOFig1}}
\end{figure}

The 8 T and 14 T spectra were obtained in a superconducting magnet; the 30 T spectra in a high homogeneity (30 ppm per mm DSV) resistive magnet at the National High Magnetic Field Laboratory.  [The $^{17}$O NMR signal cannot be independently studied between $\sim$15 T-29 T, due to overlap with the $^{139}$La$_{n=1}$ satellite.]  The applied magnetic field was oriented along the crystalline c-axis, within $\sim$2$^{\circ}$, by checking in situ the $^{139}$La$_{n=1}$ satellite position and the central transition (CT), n=0, $^{139}$La lineshape which becomes single Gaussian for H$_{0}\|$c. The unwanted $^{17}$O signal from the oxygen in La-O planes (apical oxygen, $^{17}$O$_{A}$) can be distinguished from the signal of the oxygen atoms in the Cu-O planes (planar oxygen, $^{17}$O$_{P}$) using fast pulse repetition \cite{ChenNatPhys3} to suppress the $^{17}$O$_{A}$ spectra.  This is possible because the spin-lattice relaxation rate of $^{17}$O$_{P}$ is an order of magnitude greater than $^{17}$O$_{A}$ \cite{WalstedtPRL72}.  Above 100 K the center transitions of the $^{17}$O$_{A}$ and $^{17}$O$_{P}$ have sufficiently different Knight shifts that the suppression of the apical signal can be monitored easily. For example, the remanent of the suppressed $^{17}$O$_{A}$ CT is denoted by a dot in Fig.~\ref{fig:LSCuOFig1}.  Below 100 K the apical and planar CT's overlap, so we monitor the total signal intensity and conclude that the $^{17}$O$_{A}$ signal remains negligible, consistent with previous report on La$_{1.85}$Sr$_{0.15}$CuO$_{4}$~\cite{shiftREF}.

Figure~\ref{fig:LSCuOFig1}(a-c) show CT's at 200 K for 8 T, 14 T, and 30 T.  The asymmetric lineshape at low magnetic fields is in agreement with previous reports \cite{HaasePhysC341, HaaseJoS15}.  However, the 30 T CT resolves a double peak structure not seen at lower fields.  The lineshape of the CT gives a measure of the internal magnetic field distribution and is independent of the EFG distribution for $\nu_{ref}\gg \nu_{C}$. The Knight shift $^{17}$K, i.e. the average internal magnetic field, is represented by the CT peak position, H$_{CT}$, and is defined with respect to the low temperature (4 K) value in the superconducting state \cite{shiftREF}: $^{17}$K($\%$)=[H$_{CT}$(T=4 K)-H$_{CT}$(T)]/[H$_{CT}$(T=4K)]. Figure~\ref{fig:LSCuOFig1}(d-f) show the temperature evolution of the entire NMR spectrum at 30 T. Figure~\ref{fig:LSCuOFig1}(f) shows the field-swept spectra at all three fields at 30 K where the asymmetry/double peak structure, and field-dependent broadening seen at 185 K have disappeared leaving symmetric spectra that are identical despite the large change in applied field.

\begin{figure}[t]
\begin{center}
\includegraphics[width=8cm]{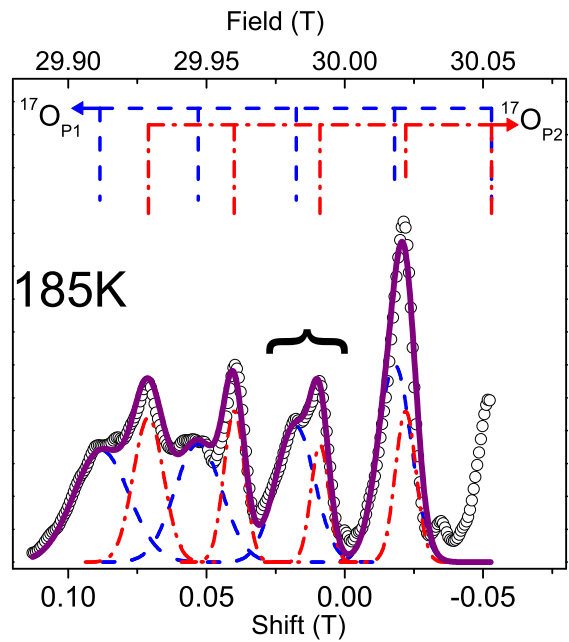}
\end{center}
\caption{Fit to 185 K $^{17}$O$_{P}$ NMR spectrum using two sets of Gaussians. Data are open circles; fit is the solid line. Fit spectra for the $^{17}$O$_{P1}$ site and $^{17}$O$_{P2}$ sites are denoted by blue dash-dot and red dashed lines, respectively. The bracket denotes the central transition.
\label{fig:LSCuOFig2}}
\end{figure}

Our most striking result is the double-peaked shape of the 30 T spectra arising from two oxygen sites in the copper oxygen plane ($^{17}$O$_{P1}$ and $^{17}$O$_{P2}$) with distinct local magnetic field and local electric field gradient distributions, even though all planar oxygen occupy equivalent sites in the orthorhombic lattice \cite{OxyBuckNote}. Figure \ref{fig:LSCuOFig2} shows the multiple-Gaussian fit to the spectrum at 185 K.  The data are well fit by two distinct quintets of quadrupolar lines~\cite{ZeroFieldTcNote, PulsesNote}, resolvable due to the large applied magnetic field. [Note that the lack of mirror symmetry of the spectrum around the CT arises from the combined effects of the Knight shift and quadrupolar splitting that sum to give a large total shift on the low-field side of the spectrum, yet partially cancel each other on the high-field side of the spectrum. See ref. \cite{HaasePhysC341}] $^{17}$O$_{P1}$ is characterized by a larger Knight shift and larger linewidth than $^{17}$O$_{P2}$. To obtain a good fit, yet minimize the number of fit parameters, the following constraints were imposed at each temperature: $^{17}$O$_{P1}$ n=$\pm$1 areas are equivalent; $^{17}$O$_{P2}$ n=$\pm$1 linewidths and amplitudes are equal. The same constraints were imposed on the n=$\pm$2 transitions.

The temperature dependence of the Knight shift of the $^{17}$O$_{P1}$ and $^{17}$O$_{P2}$ sites (Fig. 3) both decrease dramatically with temperature.  The $^{17}$O$_{P1}$ and $^{17}$O$_{P2}$ lines merge below 30 K and Knight shifts become temperature-independent below T$\sim$30 K and T$\sim$60 K, respectively.

Figure~\ref{fig:LSCuOFig4} illustrates a physically important difference between the $^{17}$O$_{P1}$ and $^{17}$O$_{P2}$ sites.  The CT linewidth (Fig 4a) measures the distribution of internal magnetic fields and is defined here as FWHM($\%$)= FWHM/H$_{0}$, where H$_{0}$ is the applied magnetic field.  The $^{17}$O$_{P1}$ FWHM is large and exhibits an increase with decreasing temperature. Together with the magnetic field dependence between 8 T and 30 T, this indicates either ferromagnetism or antiferromagnetism on the $^{17}$O$_{P1}$ sites. In contrast, the temperature and magnetic field independence of the smaller $^{17}$O$_{P2}$ FWHM indicates paramagnetism on the $^{17}$O$_{P2}$ sites.  Quadrupolar splitting in La$_{2-x}$Sr$_{x}$CuO$_{4}$ (Fig. 4b) primarily reflects the EFG from onsite hole density within the 2p-orbitals~\cite{HaasePRB69}.  For orthorhombic La$_{2-x}$Sr$_{x}$CuO$_{4}$, the EFG scales linearly with the hole density \cite{ZhengJPSJ64} another difference between the $^{17}$O$_{P1}$ and $^{17}$O$_{P2}$ sites.

\begin{figure}[t]
\begin{center}
\includegraphics[width=8cm]{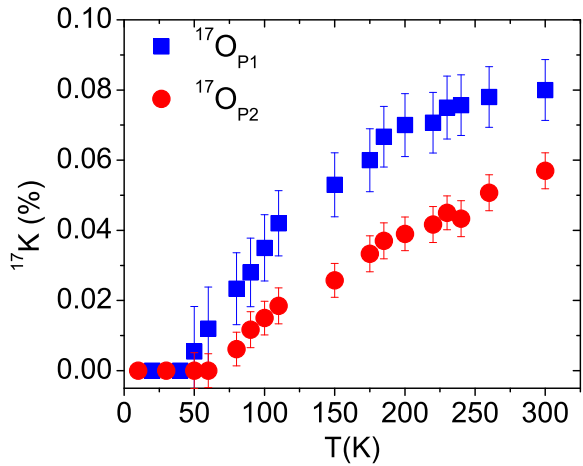}
\end{center}
\caption{$^{17}$O$_{P1}$ (blue squares) and $^{17}$O$_{P2}$ (red circles) spin shifts measured on the central transition (n=0) of the NMR spectra.  Shift is plotted relative to the low temperature value in the superconducting state, $\sim$4 K.  The most striking feature of these data is that the spin shift for  $^{17}$O$_{P2}$ drops to zero near $\sim$60 K well above T$_{C}$(H=30 T)$\sim$4 K.  The error bars are the FWHM/10 for comparison.
\label{fig:LSCuOFig3}}
\end{figure}

We conclude that Figures 1-4 clearly evidence two planar oxygen sites in underdoped La$_{2-x}$Sr$_{x}$CuO$_{4}$ with very different magnetic and charge environments. The 3:1 ratio of areas under the two distinct spectra provides an estimate of the relative number of $^{17}$O sites in each local environment.  At the lowest temperatures, roughly 75$\%$ of the planar oxygens are the $^{17}$O$_{P1}$ type and 25$\%$ are $^{17}$O$_{P2}$. We note that 19$\%$ of the planar oxygens have a nearest neighbor (nn) Sr dopant for our La$_{1.885}$Sr$_{0.115}$CuO$_4$ sample, assuming random substitution of Sr for La. However, although Sr dopants are known to shift the NMR spectrum of the nn-Cu atoms, which produces additional NMR lines, the temperature dependence of these two distinct Cu NMR spectra is the same \cite{HammelPRB57, SingerPRL2002}.  This is in sharp contrast to the distinct behaviors we report from the two Sr NMR spectra.

\begin{figure}[t]
\begin{center}
\includegraphics[width=8cm]{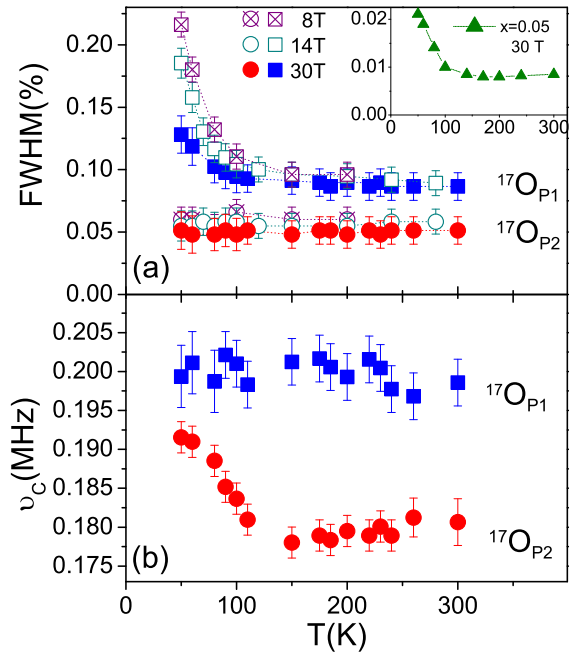}
\end{center}
\caption{Contrasting temperature dependence of NMR spectra from the $^{17}$O$_{P1}$ (squares) and $^{17}$O$_{P2}$ (circles) planar oxygen sites: (a) the central transition (n=0) FWHM/H$_{0}$ ($\%$) field dependence, where H$_{0}$ is an applied field of 8 T, 14 T, and 30 T. Inset shows behavior similar to $^{17}$O$_{P1}$ in a very underdoped, non-superconducting sample.  (b) quadrupolar splitting, $\nu_{c}$, at 30 T.
\label{fig:LSCuOFig4}}
\end{figure}

A synthesis of the data leads to the conclusion that the $^{17}$O$_{P1}$ site is characterized by (A) paramagnetism arising from delocalized holes above $\sim$100 K that (B) exhibit incommensurate antiferromagnetism at low temperatures.  The evidence for these conclusions is as follows:  (A) The $^{17}$O$_{P1}$ data in Fig.\ref{fig:LSCuOFig4}a at 8 T, 14 T and 30 T superimpose (i.e. there is a linear magnetic field dependence of the linewidth) in the temperature range 300 K to $\sim$100 K. This behavior is consistent with paramagnetism above $\sim$100 K. (B) The famous 'wipe-out' of the Cu NMR signal at T $\sim$70 K is commonly interpreted as arising from the freezing of antiferromagnetic correlations due to the localization of charge and spin configurations in a glassy magnetic state \cite{HuntPRL82, CurroPRL85, HammelPRB57, HaasePhysC341}.  Line broadening for both Cu and O lines results from the inhomogeneous distribution of internal magnetic fields in the glassy antiferromagnetic state \cite{HaasePhysC341}. The same physics gives rise to a vanishing Knight shift \cite{CuAFMcancelNote}. Our interpretation of the $^{17}$O$_{P1}$ site is consistent with recent neutron studies that reveal antiferromagnetic order at T$\sim$30 K in La$_{2-x}$Sr$_{x}$CuO$_{4}$ for x=0.10 induced by an applied magnetic field \cite{LakeNAT415}. Finally, we see similar line broadening in heavily underdoped non-superconducting La$_{2-x}$Sr$_{x}$CuO$_{4}$ with Sr concentration, x=0.05 (inset to Fig.\ref{fig:LSCuOFig4}a), a doping level that is known to be an insulating glassy antiferromagnet \cite{MatsudaPRB62}.

The $^{17}$O$_{P2}$ site, discovered here by application of very high magnetic fields, behaves very differently: it exhibits narrower resonances, a smaller spin shift and a smaller quadrupolar splitting than the $^{17}$O$_{P1}$ site.  We conclude that the $^{17}$O$_{P2}$ sites arise from (A) delocalized holes that remain metallic over our temperature range and (B) experience increased spin-singlet pairing as temperatures decrease from 300 K. The evidence for these points is as follows: (A) The temperature independence and narrower linewidths of all $^{17}$O$_{P2}$ lines (Fig.\ref{fig:LSCuOFig2}) indicates a more homogeneous distribution of hole densities, as expected from delocalized holes.  In particular the CT linewidth for $^{17}$O$_{P2}$ sites does not exhibit the magnetic broadening at low temperatures that is seen for $^{17}$O$_{P1}$ sites and the x=0.05 sample. (B) The quadrupolar splitting is constant between $\sim$100 K - 300 K indicating a fixed hole density over this temperature range. However, in Fig.\ref{fig:LSCuOFig3}, the $^{17}$O$_{P2}$ spin shift drops to zero despite this fixed hole density, indicating the spin susceptibility of the delocalized holes is zero below $\sim$60 K. In a metal, this can occur from the formation of singlet pairs.  If we accept this picture, then the zero Knight shift would indicate complete singlet pair formation on the $^{17}$O$_{P2}$ sites below $\sim$60 K, a temperature well above the superconducting transition temperature that has been suppressed to $\sim$4 K by the 30 T applied magnetic field. In fact, the temperature for singlet pair formation on the $^{17}$O$_{P2}$ sites is twice the 30 K superconducting transition temperature observed in zero magnetic field.

In conclusion, we present strong evidence for two different oxygen sites in the copper-oxygen plane of underdoped La$_{2-x}$Sr$_{x}$CuO$_{4}$ for x=0.115. Roughly 75$\%$ of the planar oxygen sites (the $^{17}$O$_{P1}$ sites) evidence local antiferromagnetism and the remaining (the $^{17}$O$_{P2}$ sites) evidence delocalized holes and spin-singlet formation. It is interesting that the NMR evidence for spin-singlet formation occurs in the same temperature range in which a finite Nernst signal has been ascribed to superconducting fluctuations \cite{XuNAT406}. Finally, we note that evidence for phase separation on oxygen sites in the copper-oxygen plane in YBCO is absent in $^{89}$Y NMR at 7.5 T \cite{BobroffPRL89} but has been claimed to exist from $^{17}$O NMR at 8.3 T \cite{HaaseJOS75}.  A very important future experiment would be to look for phase separation in YBCO using NMR at 30 teslas.

\bibliography{LSCuOall}

\begin{thebibliography}{33}
\bibitem [{Oxy()}]{OxyBuckNote}%
  \BibitemOpen
  \href@noop {} {}\bibinfo {note} {The orthorhombic phase in LSCO is created by
  the buckling of the Cu-O plane, which shifts each planar oxygen atom either
  up or down without changing the original bond lengths. This buckling does not
  create two distinct planar oxygen sites with regard to NMR.
  \cite{JorgensenPRL58}.}\BibitemShut {Stop}%
\bibitem [{Zer()}]{ZeroFieldTcNote}%
  \BibitemOpen
  \href@noop {} {}\bibinfo {note} {We note that 30 K, the temperature below
  which the spectra become symmetric, is very close to the superconducting
  transition temperature in zero magnetic field. At 30 T, T$_C$ has been
  suppressed to $\sim$4 K.}\BibitemShut {Stop}%
\bibitem [{Pul()}]{PulsesNote}%
  \BibitemOpen
  \href@noop {} {}\bibinfo {note} {Below 80 K the RF pulses are optimized for
  the CT. Above 80 K they are optimized on the n=±1 transitions.}\BibitemShut
  {Stop}%
\bibitem [{CuA()}]{CuAFMcancelNote}%
  \BibitemOpen
  \href@noop {} {}\bibinfo {note} {In the antiferromagnetic state, the Knight
  shift on the planar oxygen sites is zero due to the cancellation of the local
  internal magnetic field from the two neighboring, anti-aligned Cu moments.
  The same phenomenon occurs in La$_{1-x}$Sr$_{x}$CoO$_{3}$: the slowing of the
  Co spin fluctuations wipes out the $^{59}$Co NMR signal, and the $^{139}$La
  NMR lines exhibit a concomitant linewidth broadening
  ~\cite{RXSmithPRB78}.}\BibitemShut {Stop}%
\end{thebibliography}
\end{document}